\documentclass[11pt]{article}
\usepackage[margin=1in]{geometry}
\usepackage{amsmath}
\usepackage{amssymb}
\usepackage{enumitem}
\usepackage{comment}
\usepackage{relsize}
\usepackage{graphicx}
\usepackage{hyperref}
\usepackage{svg}
\usepackage[export]{adjustbox}
\usepackage{float}  % Needed for [H] figure placement
\usepackage{titling}
\usepackage{multicol}
\usepackage{listings}
\usepackage{xcolor}
\usepackage{authblk}
\usepackage{caption}
\newcommand{\reverted}{\rotatebox[origin=c]{180}{$\circlearrowleft$}}

\title{Using ``Failure Costs'' to Guarantee Execution Quality in Competitive and Permissionless Order Flow Auctions}

\author{}

\date{February 14, 2025}
\begin{document}
\maketitle

\begin{center}
\begin{tabular}{c@{\quad\quad}c@{\quad\quad}c}
\begin{tabular}{@{}c@{}}
  \textbf{Alex Watts}\\[6pt]
  FastLane Labs\\
  \texttt{alex@fastlane.xyz}
\end{tabular}
&
\begin{tabular}{@{}c@{}}
  \textbf{Davide Sinesi}\\[6pt]
  FastLane Labs\\
  \texttt{davide@fastlane.xyz}
\end{tabular}
&
\begin{tabular}{@{}c@{}}
  \textbf{Jacob Greene}\\[6pt]
  FastLane Labs\\
  \texttt{jacob@fastlane.xyz}
\end{tabular}
\end{tabular}
\end{center} 

\hfill \break

% Abstract
\begin{abstract}

In the context of decentralized blockchains, accurately simulating the outcome of order flow auctions (OFAs) off-chain is challenging due to adversarial sequencing, encrypted bids, and frequent state changes. Existing approaches, such as deterministic sorting via consensus layer modifications (e.g., \href{https://www.paradigm.xyz/2024/06/priority-is-all-you-need}{MEV taxes}) (Robinson and White 2024)\cite{mev_taxes} and BRAID  (Resnick 2024)\cite{braid} or atomic execution of aggregated bids (e.g., Atlas) (Watts et al. 2024)\cite{atlas_whitepaper}, remain vulnerable in permissionless settings where limited throughput allows rational adversaries to submit "spoof" bids that block their competitors' access to execution.

We propose a new \textit{failure cost} penalty that applies only when a solution is executed but does not pay its bid or fulfill the order. Combined with an on-chain escrow system, this mechanism empowers applications to asynchronously issue their users a guaranteed minimum outcome before the execution results are finalized. It implies a direct link between blockchain throughput, censorship resistance, and the capital efficiency of auction participants (e.g., solvers), which intuitively extends to execution quality. At equilibrium, bids fully reflect the potential for price improvement between bid submission and execution, but only partially reflect the potential for price declines. This asymmetry—unbounded upside for winning bids, limited downside for failed bids, and no loss for losing bids—ultimately benefits users.
\end{abstract}
% Chapter 1
\section*{1. Background}

%\begin{multicols}{2}

\noindent Order flow auctions (OFAs) have gained prominence in blockchain ecosystems as a means to improve transaction execution by allowing third-party participants (i.e., solvers or market makers) to compete for the right to fulfill orders. Commonly, OFAs grant these exclusive fulfillment rights at an initial time $t_0$, while the actual on-chain execution of the trade occurs later at time $t_1$. Ensuring that the participant who wins at $t_0$ will indeed execute the promised trade at $t_1$ is nontrivial. Without proper enforcement, a malicious actor could submit multiple high-value bids at $t_0$ and then simply refuse to fulfill them at $t_1$, undermining the auction’s reliability. Even if basic execution simulations are performed at $t_0$, market conditions can change between $t_0$ and $t_1$ in ways that invalidate those simulations, allowing participants to exploit the gap and sidestep their commitments. 

Current OFA implementations have developed various methods to enforce execution, including stateful simulations provided by block builders, reputation-based systems, and upfront payment schemes. Although stateful simulations—designed to model future execution conditions—are conceptually sound, they are not widely available across different blockchain ecosystems and often rely on relatively centralized block builder infrastructures. Reputation-based approaches (e.g., \href{https://app.uniswap.org/whitepaper-uniswapx.pdf}{UniswapX}) (Adams et al. 2023)\cite{uniswapx} limit the number of times a participant can cancel their bids between $t_0$ and $t_1$. Exceeding a cancellation threshold leads to exclusion from the set of authorized participants. However, such reputation-based strategies restrict the auction’s permissionless nature since they limit who can participate.

As \href{https://arxiv.org/pdf/2304.04981}{discussed} by Resnick (2023)\cite{resnick2023}, another challenge is that current penalty structures do not adequately incorporate the opportunity cost of failing to execute. If the market price moves unfavorably between $t_0$ and $t_1$, a participant may find it more profitable to cancel their order, even at the risk of reputation damage, rather than honor a trade that has become disadvantageous. While this behavior might still comply with formal reputation rules, it erodes the integrity of the auction. Designing penalties that fully capture this opportunity cost is difficult. Furthermore, in a permissionless environment, excluded participants can easily generate new identities to re-enter the system, negating the intended effects of reputation-based exclusions.

\subsubsection*{1.1 Mechanism Context}
\noindent This analysis proposes a new on-chain settlement approach for OFAs. Unlike traditional models that grant exclusive execution rights at $t_0$, this framework allows multiple participants to be selected at $t_0$ and then settled at execution time, $t_1$, based on their bid values and execution outcome. Such on-chain settlement preserves the capital efficiency of reputation-based methods while achieving properties similar to those of upfront payment systems. 

This approach is well-suited to environments where trusting transaction sequencers (e.g., block builders) or relying on off-chain simulations is undesirable. We assume that bidders cannot be censored by either the auctioneer or the block builder—an outcome achievable via censorship-resistant blockchain architectures such as the \href{https://arxiv.org/pdf/2301.13321}{multiple concurrent proposer model} proposed by Fox, Pai, and Resnick (2023)\cite{mccp}, or through trusted execution environments combined with cryptographic bundling protocols (e.g., the Atlas $CallChainHash$ mechanism) (Watts et al. 2024)\cite{atlas_whitepaper}. We acknowledge finite blockspace constraints at $t_1$, which naturally limit the number of contending participants. Although our primary focus is the Atlas OFA framework, these principles apply broadly to any on-chain implementation of OFAs that defer winner selection until execution time and allow atomic execution of multiple bids. \\

\subsubsection*{1.2 Atlas}

\noindent Unlike many infrastructure-driven solutions that depend on the accuracy of stateful simulations and the power to exclude conflicting transactions, Atlas is an application-specific sequencer that pairs users and solvers on-chain at execution time. As described in Section 3.3 of the Atlas \href{https://github.com/FastLane-Labs/atlas_whitepaper/blob/main/Atlas_Whitepaper.pdf}{white paper}, (Watts et al. 2024)\cite{atlas_whitepaper}, each Atlas transaction includes a user’s $UserOperation$ followed by multiple $SolverOperation$s. These solver bids are typically aggregated without pre-execution simulations, sorted optimistically by bid amount, and submitted to the blockchain as a single, large transaction. Once the transaction is included in a block, each $SolverOperation$ is executed in the sequence of their bid amounts until a solver successfully pays its bid and fulfills the user's order.

This direct, on-chain approach removes non-performing solvers even when simulations are unreliable. However, it also ties the discovery of counterparties to the blockchain’s throughput constraints (e.g., gas limits) due to the resources consumed by each $SolverOperation$ during execution. \\

\subsubsection*{1.3 Atlas Structure}

\noindent An Atlas transaction is composed of $UserOperations$ ($o_i \in O_{user}$) and $SolverOperations$ ($o_i \in O_{solver}$). Let \(G(t)\) be the gas limit of the transaction. The number of $SolverOperations$ is limited by the total gas limit and the gas consumed by user operations. Let \(g(o_i)\) denote the gas reserved for an operation \(o_i\), and \(g^*(o_i)\) the actual gas consumed. We have:
\[
\sum_{o_i \in O_{solver}} g(o_i) \;\leq\; G(t) \;-\; \sum_{o_i \in O_{user}} g^{*}(o_i)
\]
Maintaining this limit ensures that the gas needed to execute $SolverOperations$ does not exceed the remaining available gas after executing the $UserOperations$. The total gas available to all $SolverOperations$, \(\Gamma\), is:
\[
\Gamma = G(t) - \sum_{o_i \notin O_{solver}} g(o_i)
\]

$SolverOperations$ are sorted in descending order by their bid amount $b(o_i)$ and executed in sequence until one succeeds. To succeed, a $SolverOperation$ must pay its bid to the designated beneficiary and cover its own execution gas costs (minus the costs of any previously failed $SolverOperations$). If it cannot, the operation reverts but still pays for its consumed gas.

To disincentivize double-spend attacks, each solver can only submit one $SolverOperation$ per block and must escrow at least $\phi_t \cdot g(o)$ of collateral, where $\phi_t$ is the gas price of the transaction $t$.

\subsubsection*{1.4 The Censorship Vector}
\noindent In a permissionless setting, a malicious solver $i'$ can submit "spoof" bids designed purely to consume the available gas limit, $\Gamma$, rather than to win. By doing so, solver $i'$ incurs a cost proportional to the gas consumed ($\phi_t \Gamma$, where $\phi_t$ is the gas price) but gains access to any value (e.g.  \href{https://arxiv.org/pdf/1904.05234}{MEV}) (Phil Daian et al. 2019) \cite{mev_arxiv1904} $v_{i'}$ left by the user’s operation. Note that even if all solutions fail, the user’s operation might still create value that is then open to extraction. Since all other solvers’ solutions are effectively blocked from executing, the app’s designated beneficiary - typically the user - is never compensated. When the value $v_{i'}$ exceeds the cost of censorship $\phi_t \Gamma$, it becomes rational for solver $i'$ to engage in this censorship strategy.

% Chapter 2
\section*{2. Proposed Mechanism}

\noindent We introduce a new cost, referred to as the \textit{failure cost}, which applies only when a $SolverOperation$ attempts execution but fails to pay its full bid or fulfill the user's order. To calculate this cost, take the difference between the bid of the failing solver and the bid of the successful solver (if any), then multiply that difference by the fraction of the total gas allocated to the operation of the failing solver. Unlike bids in an all-pay auction, the failure cost is incurred only by a solver who secured the opportunity to execute and consumed some of the blockchain's throughput resources but did not meet their bid commitment or fulfill the user's order.

%\end{multicols}

\subsubsection*{2.1 The \textit{Failure Cost} Mechanism}
\noindent The formula for the \textit{failure cost} \( c_{fail}(o_i) \) of a $SolverOperation$ \( o_i \) is dependent on whether the array of $SolverOperation$s includes another solver's successful operation, denoted as \( b(o_j^{\checkmark}) \). Here, \( b(o_j^{\checkmark}) \) represents a \emph{successful solver operation}—that is, an operation that executes and manages to pay its bid.

\begin{equation}
\label{eq:failurecost}
c_{fail}(o_i) = 
\begin{cases} 
\left( b(o_i) - b(o_j^{{\checkmark}}) \right) \frac{g(o_i)}{\Gamma} & \text{Solver } j \text{ Succeeded, } \ b(o_j) < b(o_i)  \\
b(o_i) \frac{g(o_i)}{\Gamma} & \text{No Solver Succeeded}
\end{cases}
\end{equation}

\noindent  By combining the \textit{failure cost} with actual the gas units consumed during execution $g^*(\cdot)$ and the gas price $\phi_t$, we can describe the payoff function $V_i$ for solver $i$:

\begin{equation}
V_i = 
\begin{cases} 
v_i - b(o_i) - \phi_t \left(G^*(t) - \sum_{O^{\reverted}} g^*(o) \right) & \text{Solver } i \text{ Succeeded} \\
- \left( b(o_i) - b(o_j^{\checkmark}) \right) \frac{g(o_i)}{\Gamma} - \phi_t \cdot g^*(o_i) & \text{Solver } i \text{ Failed, Solver } j \text{ Succeeded, } \ b(o_j)<b(o_i)  \\
-b(o_i) \frac{g(o_i)}{\Gamma} - \phi_t \cdot g^*(o_i) &  \text{Solver } i \text{ Failed, No Solver Succeeded} \\
0 & \text{Solver } j \text{ Succeeded, } \ b(o_j)> b(o_i)
\end{cases}
\end{equation}

\noindent Where \(G^*(t)\) is the total gas used in the transaction \(t\) and \(O^{\circlearrowleft}\) denotes the set of operations that reverted. \\

\noindent The payoff for the auction beneficiary \( i^* \), denoted by \( W \), is defined as follows:

\begin{equation}
W_{i^*} = 
\begin{cases} 
b(o_i^{\checkmark}) & \text{First Solver Succeeded} \\
b(o_i^{\checkmark}) + \sum_{o_j \in O^{\reverted}} \left(b(o_j) - b(o_i^{\checkmark})\right) \frac{g(o_j)}{\Gamma} & \text{Mix} \\
\sum_{o_j \in O^{\reverted}} b(o_j) \frac{g(o_j)}{\Gamma} & \text{All Solvers Failed}
\end{cases}
\label{eq:userpayout} % You can use this to reference later
\end{equation}

\begin{figure}[h]
    \centering
    \includegraphics[width=1.0\linewidth, trim={20 85 20 125}, clip]{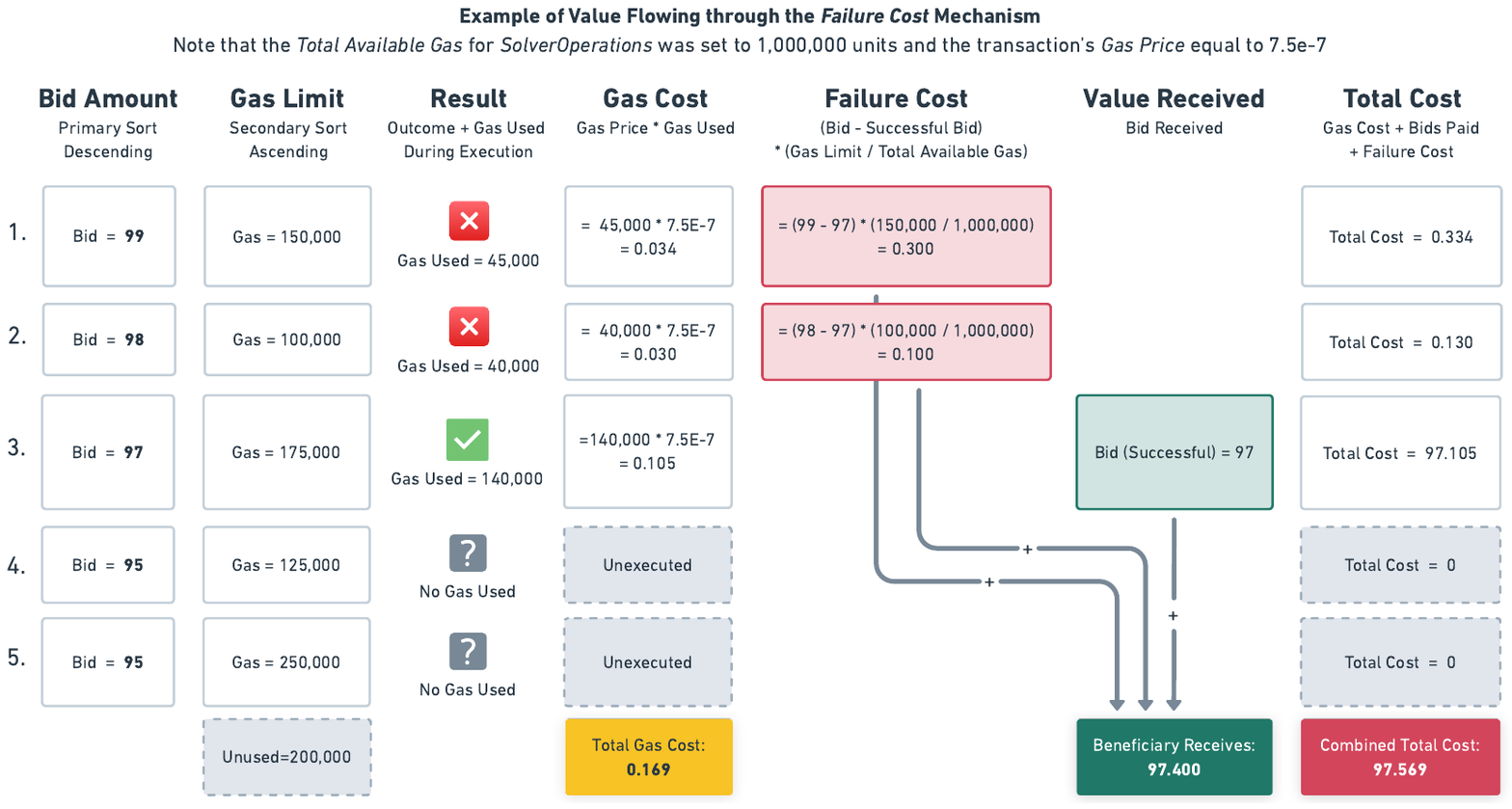} 
    \caption{An example of value flowing through the \textit{failure cost} mechanism, with the total gas available $\Gamma$ for $SolverOperations$ equal to $1,000,000$ and the gas price $\phi_t$ equal to $7.5e-7$.} 
    \label{fig:failure_example} 
\end{figure} 

% \begin{multicols}{2}

\subsubsection*{2.2 The Guaranteed Minimum Outcome}

An essential property of the \textit{failure cost} mechanism is that it enables the auctioneer in a competitive order flow auction to guarantee the beneficiary a predictable minimum payout—even in the worst-case scenario where all solvers fail. Specifically, the worst-case outcome in a competitive auction corresponds to receiving a gas-weighted average of all the bids.

\[
\sum_{i=1}^{N} \left( b_i \times \frac{g(o_i)}{\Gamma} \right) = W_{i}^{\text{worst case}}
\]

\noindent Here, \( N \) denotes the finite set of solvers participating in the auction.

\subsubsection*{2.3 Escrow and Engineering Considerations}

\noindent Atlas is designed for blockchain settings where simulations that predict order fulfillment and bid payment can be unreliable. This unreliability may arise because block times are too short relative to network latency, transactions can be \href{http://docs.monad.xyz/monad-arch/consensus/asynchronous-execution}{asynchronously sequenced by consensus} before execution (Monad 2024)\cite{monad_deferred_execution}, \href{https://www.paradigm.xyz/2020/08/ethereum-is-a-dark-forest}{adversarial entities} can insert conflicting transactions between simulation and inclusion (Robinson and Konstantopoulos 2020)\cite{dark_forest}, or the blockchain’s state may be subject to reorganization.

While Atlas mitigates these issues by selecting the highest-bidding solution atomically at execution time, the capital backing \textit{failure costs} remains vulnerable. For example, an adversarial solver could \href{https://arxiv.org/pdf/2308.06513}{front-run} (Oz et al. 2024)\cite{front_run_arxiv2308}
the Atlas transaction with its own transaction that depletes its balance before the failure penalty is applied.

To address this, we propose extending Atlas’s gas escrow to also include bid balances. Recall that this mechanism's auctioneers are responsible allocating their auction's limited throughput resources, which they do by choosing the set of $SolverOperations$ executed on-chain - they do not select the winning solution. Since the auctioneers cannot predict which solver solutions will succeed, each solver must maintain an escrow covering their worst-case penalty $b(o_i) \frac{g(o_i)}{\Gamma}$ and their potential gas costs for all unexecuted, pending $SolverOperation$s. Formally:
\[
e_{i,a}(o_i) \ \ge \sum_{o_i \in O_{i,a}^?} \ b(o_i) \frac{g(o_i)}{\Gamma} \ + \   \phi_t \cdot g(o_i)
\]
where $a$ is the auctioneer, $O_{i,a}^?$ is the set of $SolverOperation$s from solver $i$ handled by auctioneer $a$ whose execution outcomes are unknown, and $e_{i,a}(o_i)$ is the required escrow for operation $o_i$.

The auctioneer can pre-fetch and cache each solver’s escrow balances. With local "in flight" accounting of potential expenditures and a direct mapping of escrowed balances to auctioneers, the auctioneer can sequence $SolverOperation$s without querying the blockchain or performing simulations during the live auction.

\subsubsection*{2.4 Asynchronous User Experience}

The \textit{failure cost} mechanism, combined with the “in flight” accounting system, allows applications to provide a guaranteed minimum outcome without relying on additional blockchain reads or writes between receiving a user’s order and issuing the guarantee. This greatly improves the user experience (UX):

\begin{itemize}
    \item At $t_{-2}$ or earlier, the auctioneer pre-fetches and caches each solver’s escrow balances.
    \item At $t_{-1}$, the auctioneer receives a user’s order, initiates the order flow auction and begins collecting bids from solvers.
    \item At $t_{0}$, once bidding ends, the auctioneer can immediately compute and provide the user with a guaranteed minimum outcome while simultaneously submitting the transaction to the blockchain.
    \item At $t_1$, the blockchain executes the transaction and finalizes the actual outcome.
\end{itemize}

Because no on-chain operations are needed during the auction itself, the user’s waiting time is limited only by network and auction durations. For example, if user-to-auctioneer latency is $50ms$ and the auction lasts $300ms$, the user can view their guaranteed minimum outcome within about $400ms$ of placing the order, providing a fast and responsive experience comparable to off-chain trading environments.
% Chapter 3
\section*{3. Mechanism Properties}

\noindent The mechanism’s design ensures that participants behave in ways that maximize user value, discourage censorship, and maintain overall market integrity. Specifically, we identify the following desirable properties:

\begin{enumerate}
    \item \textbf{User-oriented value maximization:}  
    The user’s payoff is highest when the top-bidding solver completes its operation successfully. Since users do not gain additional value if solvers fail, their optimal outcome is always to secure a successful, well-priced solution. This aligns user incentives with the mechanism’s fundamental objective—efficient and reliable order execution.

    \item \textbf{Incentives for subsequent solver success:}  
    If a solver fails, it becomes that solver’s interest for the next solver to succeed. Because $SolverOperation$s are executed in descending order of their bid amounts, each subsequent solver’s success reduces the failure cost borne by the previously failing solvers. In other words, when a solver fails, it does not benefit from further failures; instead, it prefers that a successor solver succeed to minimize its own penalty.

    \item \textbf{Penalties scale with resource usage:}  
    Solvers reserving more gas resources are effectively penalized more in the event of failure. As a solver’s relative gas consumption \(\frac{g(o_i)}{\Gamma}\) increases, so does its potential failure cost. This creates a natural deterrent against unnecessarily large or gas-intensive operations that would otherwise crowd out other participants.

    \item \textbf{Censorship requires comprehensive dominance:}  
    Consider a solver \(i'\) aiming to censor all other solvers. To do so, \(i'\) must (1) outbid every other solver, and (2) reserve the entire (or nearly entire) block of gas (\(\Gamma\)) allocated to $SolverOperation$s:
    \[
    b(o_{i'}) > b(o_i) \quad \forall o \in O, \quad \text{and} \quad \sum g(o_{i'}) \geq \Gamma
    \]
    Under these conditions, the cost to the censoring solver is:
    \[
    b(o_{i'}) + \phi_t \cdot g^*(o_{i'})
    \]
    which strictly exceeds the combined bids of the censored solvers. If the censoring solver values its chance of winning the auction (and thereby capturing $v_{i'}$) at least as much as it values the act of censorship, it is always more profitable in expectation to execute successfully rather than revert and censor others. This dynamic ensures that censorship, while theoretically possible, is made less attractive by the mechanism’s incentive structure. Moreover, if the censoring solver does succeed, the user’s payoff corresponds to \(b(o_{i'})\), which is strictly greater than the user’s payoff in the absence of censorship, thereby preserving user welfare.

\end{enumerate}

\subsubsection*{3.1 Escrow and Capital Efficiency}

\noindent The capital efficiency of the auction participants depends on $\Gamma$ the total gas allocated to solvers. Increasing $\Gamma$ reduces the share of total gas reserved by any single solver, thereby lowering their worst-case penalty and the \textit{failure cost}'s escrow requirements. As solvers often act as counterparties or market makers, the intuition is that blockchains with higher throughput are more capital efficient.

\begin{comment}
\begin{figure}[h]
        \includesvg[width=0.5\linewidth]{figure_5_fix}
        \caption{Total Gas Available and Capital Efficiency}
        \label{fig:total_gas_and_capital_efficiency}
\end{figure}
\end{comment}

Conversely, decreasing $\Gamma$ forces solvers to escrow more capital, reducing efficiency. However, higher $\Gamma$ not only enhances efficiency but also strengthens censorship resistance, as solvers have more room to operate, increasing competition. The gap between required escrow and failure cost remains fixed at $\phi_t \times g(o_i)$, a term independent of $\Gamma$.

\begin{comment}
\begin{figure}[H]
    \centering
    \begin{minipage}[b]{0.45\linewidth}
        \vspace{30pt}
        \includesvg[width=\linewidth]{figure_5_fix}
        \caption{Total Gas Available and Capital Efficiency}
        \label{fig:total_gas_and_capital_efficiency}
    \end{minipage}
    \hfill
    \begin{minipage}[b]{0.45\linewidth}
        \includesvg[width=\linewidth]{figure_4_fix_v2}
        \caption{Required Escrow and \textit{Failure Costs}}
        \label{fig:required_escrow_and_failure_costs}
    \end{minipage}
\end{figure}
\begin{figure}[H]
    \centering
    \begin{minipage}[b]{0.45\linewidth}
        \raisebox{-100mm}{\includesvg[width=\linewidth]{figure_5_fix}}
        \caption{Total Gas Available and Capital Efficiency}
        \label{fig:total_gas_and_capital_efficiency}
    \end{minipage}
    \hfill
    \begin{minipage}[b]{0.45\linewidth}
        \includesvg[width=\linewidth]{figure_4_fix_v2}
        \caption{Required Escrow and \textit{Failure Costs}}
        \label{fig:required_escrow_and_failure_costs}
    \end{minipage}
\end{figure}
\end{comment}

\begin{figure}[H]
    \centering
    %\includesvg[width=\linewidth]{figures/figure_2}
    \includegraphics[width=\linewidth]{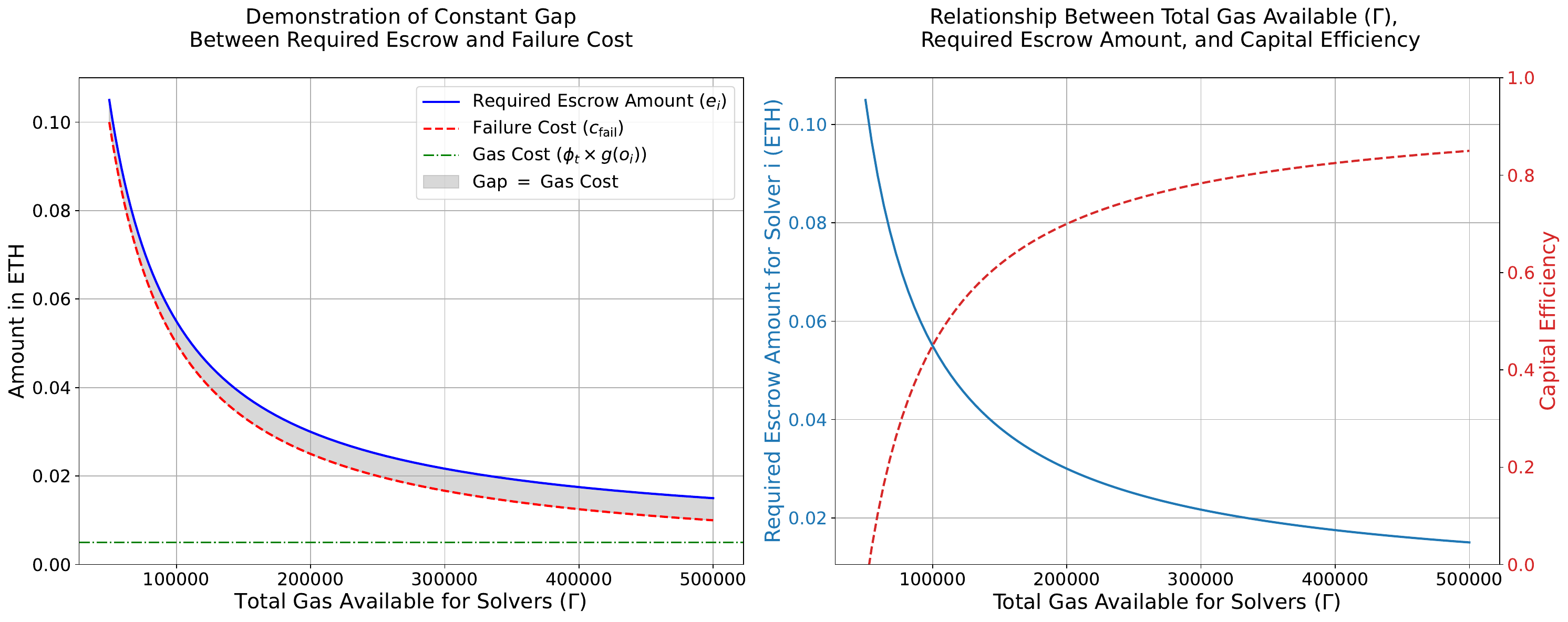}
    \caption{Combined Plots of Gas Availability and Costs}
    \label{fig:combined_charts}
\end{figure}

\begin{comment}
\begin{figure}[h]
    \centering
    \includesvg[width=0.5 \linewidth]{figure_4_fix_v2}
    \caption{Required Escrow and \textit{Failure Costs}}
    \label{fig:required_escrow_and_failure_costs}
\end{figure}
\end{comment}

\noindent In addition to directly lowering the maximum \textit{failure cost} \( c_{fail}(o_i) \) for each $SolverOperation$ as throughput \(\Gamma\) grows, an increase in \(\Gamma\) also indirectly reduces the expected failure cost \(\mathbb{E}[c_{fail}(o_i)]\). This secondary effect occurs because larger \(\Gamma\) values allow more $SolverOperation$s to be included, increasing the likelihood that at least one succeeds and thus diminishes the expected penalty faced by any failing solver. \\

\noindent \textbf{Example:}  
Suppose a $SolverOperation$ \(o_i\) uses \(g(o_i)=100{,}000\) gas and bids \(b_i(v_i) = \$100\). If \(\Gamma_1 = 1{,}000{,}000\), then:
\[
\max[c_{fail}(o_i)] = \frac{100{,}000}{1{,}000{,}000} \times 100 = \$10.
\]
If the gas usage is similar across solvers, we might expect around 10 $SolverOperation$s in the transaction. As a median bid, if \(o_i\) fails, about 5 other operations remain to succeed and mitigate \(c_{fail}(o_i)\).

Now consider \(\Gamma_2 = 10{,}000{,}000\):
\[
\max[c_{fail}(o_i)] = \frac{100{,}000}{10{,}000{,}000} \times 100 = \$1.
\]
With this higher throughput, we might now expect around 100 $SolverOperation$s. If \(o_i\) fails as a median bid, about 50 other operations can succeed and further reduce \(\mathbb{E}[c_{fail}(o_i)]\).

Thus, two factors lower \(\mathbb{E}[c_{fail}(o_i)]\) when \(\Gamma\) increases:
\begin{enumerate}
    \item The maximum possible failure cost per operation decreases proportionally to \(\Gamma\).
    \item The probability that at least one other solver succeeds (and thereby reduces the penalty) increases as more $SolverOperation$s fit into a single auction.
\end{enumerate}

Define \( N^* = \mathbf{card}\{ o_j : b(o_j) < b(o_i) \} \) as the number of strictly lower-bidding $SolverOperation$s than \(o_i\). As \(\Gamma \to \infty\):
\[
\mathbb{E}[N^*] \to \infty,
\]
\[
P\left( \exists \, o_j \in O^{\checkmark} \mid o_i \in O^{\reverted} \right) \to 1,
\]
\[
b(o_i) - \mathbb{E}[b(o_j) \mid o_i \in O^{\reverted}] \to 0.
\]

Since the \textit{failure cost} depends on the difference between \(b(o_i)\) and the successful bids \(b(o_j^{\checkmark})\), and this difference approaches zero, the expected failure cost also approaches zero. Formally:
\[
\lim_{\Gamma \to \infty} \mathbb{E}[c_{fail}(o_i)] = 0.
\]

\subsubsection*{3.2 Censorship Resistance}

\noindent The \hyperref[eq:costofcensorship]{cost-of-censorship expression} derived earlier over-simplifies the problem and assumes that the censoring solver must reserve all available gas. In practice, a censoring solver need only reserve:
\[
\Gamma - \min_O \{g(o)\}
\]
where \(\min_O \{g(o)\}\) is the smallest gas allocation among the other solvers. By reserving just enough gas to exclude all other solvers, the censoring solver reduces its cost of censorship by:
\[
\frac{\min_{O}(g(o))}{\Gamma}
\]

This adjustment is critical for understanding how to choose an optimal \(\Gamma\) and how to relate blockchain throughput directly to censorship resistance. To analyze this further, define:
\[
\Gamma_{i'}' = \Gamma - \min_{O} \{ g(o_i) : i \neq i' \} 
\]
\[
B_{i'} = \max_{O} \{ b(o_i) : i \neq i' \} \\ 
\]
Here, \(\Gamma_{i'}'\) represents the effective gas limit after accounting for the smallest $SolverOperation$’s gas usage, and \(B\) represents the largest competing bid. The censorship threshold can now be expressed as:
\begin{equation}
\text{Censorship Resistance} = \Gamma_{i'}' \left( \phi_t + \frac{B}{\Gamma} \right) - v_{i'}
\label{eq:costofcensorship}
\end{equation}

\begin{figure}[h]
    \centering
    %\includesvg[width=0.5\linewidth]{figures/figure_3}
    \includegraphics[width=0.6\linewidth]{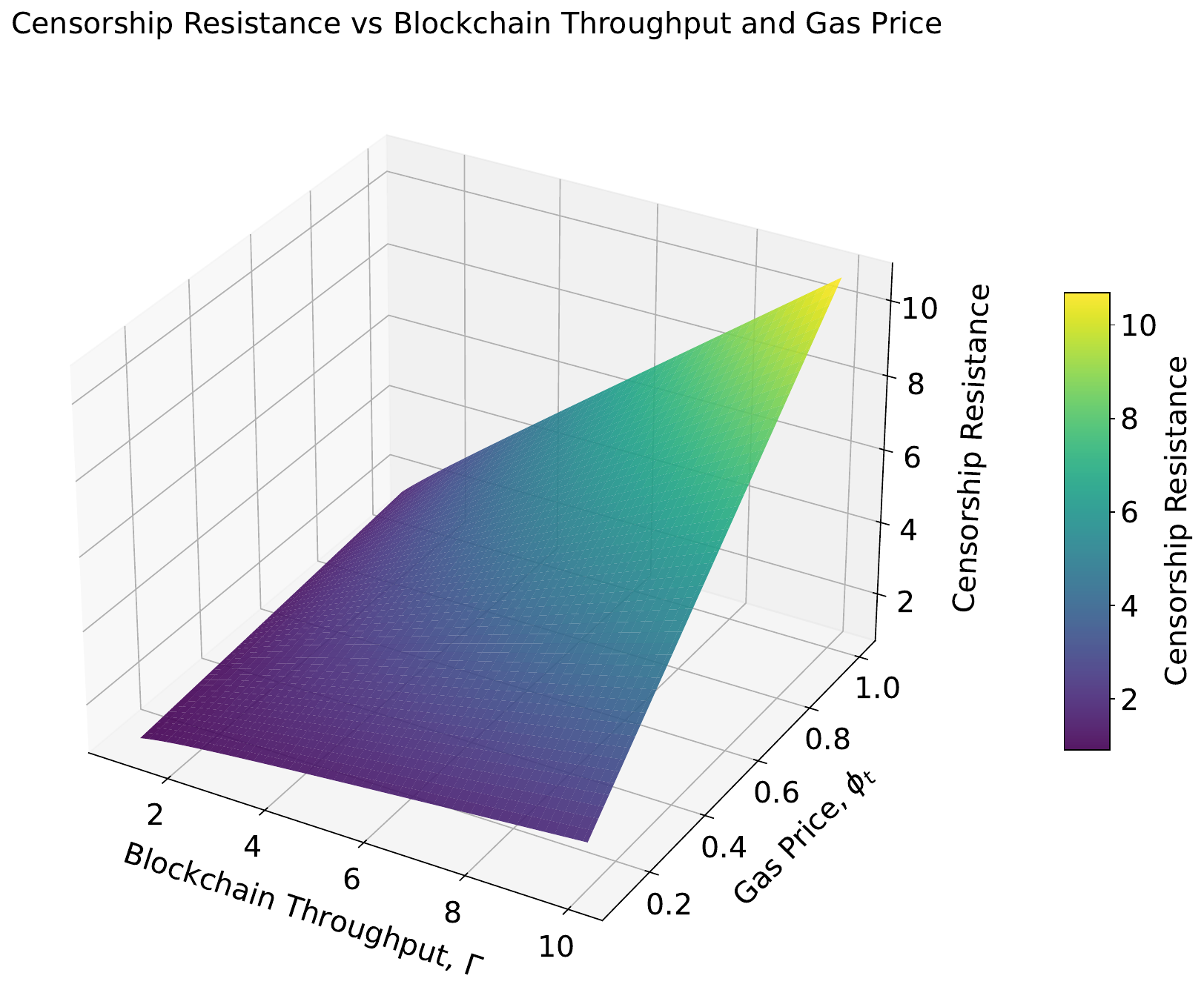}
    \caption{Relationship between Censorship Resistance, Gamma, and Gas Price}
    \label{fig:censorship_resistance_2}
\end{figure}

This refined formulation provides a more nuanced perspective on how increasing blockchain throughput (\(\Gamma\)) enhances censorship resistance. As \(\Gamma\) grows, censorship resistance improves, making it more difficult and less profitable for adversaries to censor competing operations. 
Figure~\ref{fig:censorship_resistance_2} illustrates how variations in gas price and gas available to solvers (throughput), highlighting throughput as a crucial design lever for maintaining a robust and equitable auction environment.

\begin{comment}
\begin{figure}[H]
    \centering
    \includesvg[width=\linewidth]{figure_1_fix}
    \caption{Censorship Resistance Map}
    \label{fig:censorship_resistance}
\end{figure}
\end{comment}

% Chapter 4
\section*{4. Bidding Strategies}
Solvers choose their bids based on expectations about future values and the behavior of others. Because the final outcome depends on everyone’s decisions, no single solver can simply choose a strategy without considering the likely actions of others. The mechanism is therefore not strongly incentive compatible.

Instead, the mechanism leads to a Bayes-Nash equilibrium. Although this equilibrium ensures that no solver can profit by changing their own strategy once the equilibrium is established, it does not mean that solvers simply bid their true values. A rational solver would adjust their bid in light of the probabilities of winning, potential penalties, and anticipated moves by competitors. Strongly dominant strategies are unlikely given the complexity, uncertainties, and strategic interactions between participants.

\subsubsection*{4.1 Theoretical Framework}

\noindent Consider a setting in which solvers, unaware of their counterparts’ bids, must select a strategy that maps their private valuation \( v_i \) to an optimal bid \( b_i(v_i) \).

\vspace{1em}

\noindent To review the operational sequence:
\begin{itemize}[label=-, left=10pt, itemsep=0.5em]
    \item Let \( n \in \mathbb{N} \) denote the (finite) number of solvers participating in a given auction. Note that \( n \) is not predetermined but depends on factors such as the available throughput and the gas consumption of user operations.
    \item Each solver must escrow sufficient funds to guarantee coverage of potential failure costs.
    \item \( SolverOperations \) are executed in descending order of their bids until one operation is successfully completed.
    \item A \( SolverOperation \) is deemed successful if it pays its own bid, its associated gas costs, and fulfills the user's order by the conclusion of its execution.
    \item Any \( SolverOperation \) that is executed but fails to cover its bid, fulfill the order, or pay its gas costs is penalized via the specified \textit{failure cost} mechanism.
\end{itemize}

\noindent We consider an environment where each participant (i.e., bidder) faces an independent probability \( q \in (0,1) \) from  \href{https://arxiv.org/pdf/2304.04981}{M. Resnick (2003)} \cite{resnick2023} that their operation fails to fulfill its obligation, and there are \( n > 1 \) active bidders in a sealed-bid, first-price auction. Each bidder’s valuation of winning the auction is the commonly known quantity \( v \). Suppose we focus on a symmetric profile where all other bidders bid a common amount \( b \). We aim to determine the best response for a single bidder under this assumption---specifically, whether it is profitable to deviate from \( b \) by a small increment \(\varepsilon > 0\).

While these assumptions are strong - especially $q$ being independent between solvers - they serve as a useful benchmark. If valuations and all public signals truly remained fixed, simulations would yield deterministic outcomes, leaving no incentive for a winner to default on their promised bid. Yet, real-world conditions are more complex, and the proposed mechanism must accommodate the possibility of unexpected failures at execution time.

When all bidders, including the focal bidder $i$, submit the same bid \( b \), the expected utility is:
\[
U(b) \;=\; v\,(1 - q^n) \;-\; b.
\]
This arises because, with probability \((1 - q^n)\), the bidder's bid ``succeeds'' and they pay \( b \) to secure value \( v \), thereby obtaining a net utility of \( v - b \).

By bidding slightly more, \( b + \varepsilon \), the expected payoff changes according to the probabilities of bid success or failure. Intuitively, this is because bidder $i$ will have the highest bid in the array and will be executed first. When all players other than $i$ bid $b$, we model the utility to $i$ of bidding $b + \epsilon$:
\[
U_i\bigl(b_i =b + \varepsilon, b_{-i}=b\bigr)
\;=\;
(1 - q)\,\bigl[v - (b + \varepsilon)\bigr]
\;+\;
q\,\frac{-\,\varepsilon}{n}
\;+\;
q^n \,\frac{-\,b}{n}.
\]
Intuitively, with probability \((1 - q)\), the bidder wins and pays \( b + \varepsilon \) while receiving \( v \). In the complementary states of failure, the bidder pays only a fraction of the submitted amount, depending on how many other bidders fail simultaneously. To find the symmetric-equilibrium bid \( b \), we set up the difference
\[
\Delta(\varepsilon)
\;=\;
U\bigl(b + \varepsilon\bigr) \;-\; U(b),
\]
and examine the limit of \(\Delta(\varepsilon)\) as \(\varepsilon \to 0\). At equilibrium, we require \(
\lim_{\varepsilon \to 0} \Delta(\varepsilon) \;=\; 0,
\) so that the bidder is indifferent between bidding \( b \) and slightly deviating to \( b + \varepsilon \). 

 By equating that limit to zero and performing straightforward algebraic manipulation, we obtain the following closed-form expression for the symmetric equilibrium bid, denoted \( b^* \):
\[
b^*
\;=\;
v \,\cdot \frac{\,n\,\bigl(1 - q^{\,n-1}\bigr)}{\,n - q^{\,n-1}\!}.
\]
The factor \(\frac{n\,(1 - q^{\,n-1})}{\,n - q^{\,n-1}\!}\) reflects how the failure probability \( q \) and the number of bidders \( n \) jointly affect the bid level in this probabilistic, first-price mechanism. Note that while a bid of zero might appear to have utility at first glance by paying nothing to capture the expectation of all other bidders failing, in practical applications of this design this is a non-issue due to the existence of gas costs and the presence of $>n$ bids - only the top $n$ are included in the array.

This result, though derived under generous assumptions, provides a powerful intuition: as the number of solvers \( n \) increases, the geometric influence of \( q^n \) ensures that even substantial probabilities of failure are tempered. A higher \( n \) can offset a higher \( q \), maintaining incentives to bid more assertively than one might expect from a large failure probability alone. The resulting relationship offers a theoretically sound baseline for understanding how the scale of competition (reflected in \( n \)) and the underlying risk (reflected in \( q \)) jointly shape equilibrium bidding behavior in complex, dynamically evolving blockchain ecosystems.

\subsubsection*{4.2 Extending the Model with Discrete Time}

\noindent In the previous analysis, we assumed that private valuations and public information remained fixed from the moment bids are placed ($t_0$) until the auction settles ($t_1$). In practice, this is rarely the case. Valuations can evolve, external signals may arrive, and clever adversaries can exploit changing conditions. To capture these complexities more faithfully, we now extend our model to incorporate the potential for valuations to change over time and the strategic implications this creates.

Previously, we introduced a single probability of failure $q$, treating it as an exogenous factor, but in a truly permissionless setting we must acknowledge the presence of adversarial solvers and asymmetric information. A solver cannot alter its posted solution ex post, but it can undertake minimal-cost actions—such as toggling an “off switch” in a smart contract or rapidly reallocating un-escrowed resources—to force an ostensibly winning solution to fail. This potential for strategic manipulation complicates the relationship between posted bids and realized outcomes, particularly as time and state evolve.

As a solver’s relative bid increases, it indirectly creates more “attempts” from lower-bidding solvers to succeed and thereby mitigate its own potential penalty. In other words, the higher a solver's bid relative to its competition, the greater its expectation of the number of solutions available to execute should it fail and backstop its failure penalty. This interplay heightens the strategic complexity: the marginal incentive to bid higher depends not only on expected value and risk but also on the dynamic interaction between other solvers’ future actions and the evolving price distribution.

We define $v$ as the commonly known value of the asset at $t_0$, while at $t_1$ the asset’s price follows a normal distribution \(X \sim N(v,\sigma^2)\) with mean $v$, standard deviation $\sigma$, CDF $F_X(x)$, and PDF $f_X(x)$. Solvers must commit their bids at $t_0$ based solely on the expected distribution of $X$, yet at $t_1$ each solver privately observes its own realization of $X$—but not that of its competitors. This observational asymmetry enables a winning solver to extract positive slippage when $X > v$, while providing a cushion against losses due to the bounded failure penalty $\frac{b(v)}{n}$. 

If at the time of execution $t_1$ the solver observes $X_i < b(v_i)$, it will cancel rather than incur a larger loss. We can incorporate this concept into our utility framework by replacing $X$ in the payoff calculations with its conditional expectation given $X > b(v)$:
\[
\mathbb{E}[X \mid X > b(v)] = \int_{b(v)}^{\infty} \frac{x f_X(x)}{1 - F_X(b(v))} dx
\]

Note that the decision threshold for profitability remains firmly at $X > b(v)$, not $X-\text{penalty} > b(v)$. This is because $b(v)$ already addresses the expectation of the failure penalty, which informs the solver’s choice to cancel but does not redefine the fundamental break-even point for profitable trade execution. Substituting in this conditional expectation of $X$ and recalling the normality assumption for $X$ yields our utility function $U_i \left[ b_i(v), \ b_{-i}(v) \right]$:

\begin{align*} 
n 
\left[ 
    v (1- F_X(b(v))) 
    + \sigma f_X(b(v)) 
    - \frac{ b(v) (1 - F_X(b(v))) }{ 1 - [F_X(b(v))]^n }
\right]
\end{align*}

Unlike simpler models, no neat closed-form solution emerges. Instead, the equilibrium depends on the entire distribution of future values and on strategic cancellation options that arise as solvers observe their private realization of $X$.

\subsubsection*{4.3 Numerical Analysis }

The equilibrium bid $b^*(v)$ is found by setting the derivative of $U_i[b(v)]$ with respect to $b(v)$ to zero. Letting $z = \frac{b(v)-v}{\sigma}$ and recognizing that $\Phi(z)$ and $\phi(z)$ represent the normal CDF and PDF respectively, we derive a complex condition for equilibrium (as shown in Appendix I). 

When using numerical optimization solutions to solve for the optimal bid, we observe that $b^*(v)$ exceeds $v$, and that it increases with both $\sigma$ and $n$. This matches our expectations given the assumptions - as long as the solver sampling in $X$ is independent, additional solvers act to reduce the expectation of a failure penalty, and as the standard deviation increases, the solvers return in expectation increases with their expectation of positive slippage augmented by a lower bound on the penalty. \\

\begin{figure}[h]
    \centering
    \includegraphics[width=0.5\linewidth]{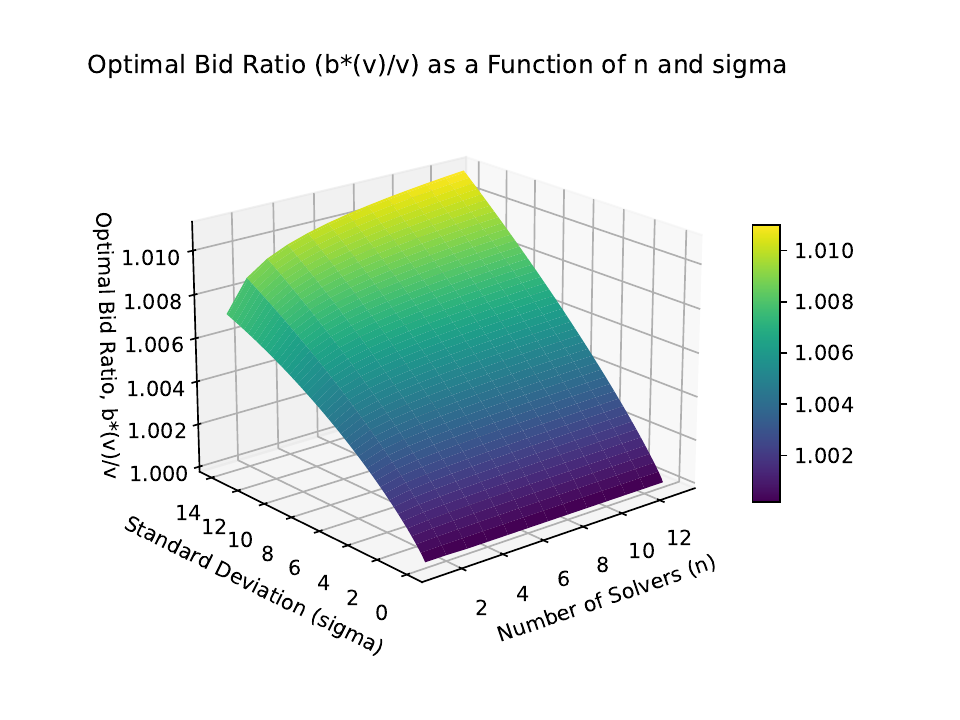}
    \caption{The optimal bid ratio, $\frac{b^*(v)}{v}$, as a function of $n$ and $\sigma$, with an initial value of $v=\$3,500$ and a range of $[\$0, \$ 12.00]$ for $\sigma$.}
    \label{fig:optimal_bid_n_sigma}
\end{figure}

\subsubsection*{4.4 Model Limitations }

It is important to note that the model proposed in this section assumes that solver valuations sampled from $X$ are independent. In reality, however, these valuations are highly covariant, and solver competition often results in congestion costs that reduce the overall welfare of participants \href{https://arxiv.org/pdf/2403.02525}{(Chitra, Kulkarni, Pai, and Diamandis 2024)} \cite{analysis_intent_markets}. 

Another important caveat is the simplifying assumption that the expectation of solver bids are rank-agnostic: $\mathbb{E}[b(v_i)] \approx \mathbb{E}[b(v_j) \,| \,b(v_j)<b(v_i)]$. If all solvers bid the same $b(v)$ then one solver's success fully offsets the failure cost of all the others. In reality, differences in the private valuations at $t_0$ would lead to a distribution of bids, which would increase the expectation of failure costs. A model treating each bid as an order statistic may lead to a more applicable framework for deriving an optimal bid formula.

While the assumptions of independent $X$ samples and rank-agnostic bid expectations are likely resulting in inflated optimal bid values, the potential for the \textit{failure cost} mechanism to unlock a path towards users internalizing a portion of the expectation of positive slippage is promising. Understanding how correlated or covarying valuations affect equilibrium bidding behavior, the distribution of failure costs, and the dynamics of censorship resistance would significantly deepen our theoretical understanding and practical guidance. Future research that incorporates these complexities would improve the model's predictive power and deepen our knowledge of the mechanism.

% Chapter 5
\section*{5. Implications \& Observations}

Beyond its primary role in enhancing censorship resistance, the \textit{failure cost} mechanism also encourages more responsible bidding behavior. Without a penalty for failed bids, solvers might submit numerous low-quality or speculative solutions (i.e., high bid, low probability of payment), consuming valuable resources. Here, the financial penalty for failure motivates solvers to ensure their solutions are valid and likely to succeed.
\subsubsection*{5.1 Permissionlessness}

The \textit{failure cost} mechanism allows solvers to participate in order flow auctions (OFAs) without relying on a reputation system. As long as they maintain the required escrow balances, their bids can be considered. This makes the OFA inherently permissionless unless the auctioneer chooses to impose additional restrictions.

While permissionless bidding may influence execution quality, achieving optimal execution is not the only goal. Fundamentally, permissionlessness aligns with the core principles of decentralized blockchains and their applications. We believe this attribute is inherently valuable, and we hope readers appreciate its importance in fostering open, accessible, and censorship-resistant ecosystems.

\subsubsection*{5.2 Blockchain Throughput}
Intuitively, a blockchain’s throughput (total gas per block) is an upper bound for an on-chain auction's throughput (total gas per OFA), which directly influences both censorship resistance and execution quality. Increasing the throughput makes censorship more resource-intensive and less economically viable, encouraging legitimate solvers to participate and submit robust solutions. A higher throughput allows more $SolverOperation$s to fit into a block, increasing competition and pushing bids upward, ultimately improving the beneficiary’s payoff. 

Reduced censorship risk and greater solver participation lead to more frequent successful executions and lower \textit{failure costs} in expectation, prompting solvers to bid more aggressively. This dynamic reduces the amount of capital each solver must escrow, which reduces barriers to entry, improves execution outcomes, and increases capital efficiency of market makers - the blockchain's primary active liquidity providers.

\subsubsection*{5.3 Applications to Cross-Chain Orders}

Beyond enhancing UX in a single-chain context, the ability to offer a guaranteed minimum outcome without on-chain state checks has potential implications for cross-chain interactions. In cross-chain swaps (also known as intent-based bridging or chain abstraction), transactions may span multiple chains, introducing delays as information and assets move through bridges.

If the guaranteed minimum outcome can be computed asynchronously—without waiting for confirmations or state changes on the destination chain—users might receive timely assurance on their eventual outcome even when assets have not yet crossed the bridge. This early assurance could reduce uncertainty, encourage participation in multi-chain strategies, and improve overall efficiency in cross-chain trades. For complex orders requiring multiple interdependent cross-chain interactions, the guaranteed minimum outcomes of each step could be calculated and used as the foundation of the following step, enabling multiple solvers to trustlessly coordinate while executing the steps in parallel.

These potential benefits highlight the need for further research. How might this asynchronous guarantee interact with various bridging protocols, liquidity conditions, or complex multi-chain topologies? Investigating these questions could uncover valuable insights into extending the mechanism’s application. 
% Conclusions
\subsubsection*{Conclusion}

In this work, we introduced and analyzed a novel on-chain settlement mechanism for order flow auctions centered around the concept of a \textit{failure cost}. This mechanism imposes a structured penalty on solvers who gain the right to execute but fail to pay their bid, ensuring that participants bear real consequences for low-quality, manipulative, or obstructive bidding tactics. By doing so, it strengthens censorship resistance, discourages spam or spoof bids, and guarantees the beneficiary a predictable minimum payout—even under worst-case conditions.

Our analysis demonstrates that the mechanism leads participants toward a Bayes-Nash equilibrium, where no single solver can profitably deviate from their chosen strategy once the equilibrium is established. While the mechanism does not ensure strict incentive compatibility or dominant strategies, it does encourage bidding behavior guided by rational, forward-looking expectations. Solvers must carefully weigh private valuations against the risk of failure, adjusting their bids to maximize their expected payoff in a competitive and uncertain environment.

In a more dynamic scenario where valuations and external information evolve between bid submission and execution, the complexity and strategic considerations deepen. Nonetheless, the key principles remain: the \textit{failure cost} mechanism aligns economic incentives, encourages meaningful engagement from solvers, and fosters an environment in which quality, reliability, and well-informed bidding become the norm. As decentralized finance and blockchain-based auctions continue to mature, these insights—rooted in equilibrium analysis, throughput considerations, and the interplay of risk, valuation, and competitive strategies—offer a robust theoretical and practical foundation for designing next-generation auction protocols that serve both solvers and end-users more effectively.

\newpage
% Acknowledgements
\subsubsection*{Acknowledgements}
We would like to thank Tom Mroz, Christian Saur, Nagu Thogiti and the FastLane team for their advice, input, and support in the creation of this paper. 

\newpage
% Bibliography

\newpage
% appendix
\appendix
\section*{Appendix I - Math}

\subsubsection*{Foundational Strategy}

Solving for the utility function: \\ \\ 

\noindent Background: bidder $-i$ bids $b$ \\
\noindent Scenario 1: 
\textit{"If I want to bid more than $b$ I will go first, so I bid $b + \epsilon$"}
\begin{align*}
   U_i(b_i=b+ \epsilon) = (1-q)[v - b - \epsilon] 
        \ + \ 
    q \frac{ - \epsilon }{n} 
        \ + \
    q^{n} \frac{ -b}{n}
\end{align*} 

\noindent Scenario 2: 
\textit{"I want to bid the same as everyone else, so I bid $b$"}

\begin{align*}
U_i(b_i=b+ \epsilon) 
&= \sum_{k=0}^{n-1} 
\Bigl[
  (1 - q^k)\cdot 0
  \;+\;
  q^k (1-q)(v-b)
  \;+\;
  q^k\,q\,\Bigl(\tfrac{-b}{n}\Bigr)
  \;+\;
  q^k\,q\,\Bigl(1 - q^{\,n-k-1}\Bigr)\Bigl(\tfrac{b}{n}\Bigr)
\Bigr]
\\[6pt]
&=\;
\sum_{k=0}^{n-1} 
\Bigl[
  q^k (1-q)(v-b)
  \;-\;
  \tfrac{b}{n}\,q^{k+1}
  \;+\;
  \tfrac{b}{n}\,q^{k+1}\Bigl(1 - q^{\,n-k-1}\Bigr)
\Bigr]
\\[6pt]
&=\;
\sum_{k=0}^{n-1} q^k (1-q)(v-b)
\;+\;
\tfrac{b}{n}\,
\sum_{k=0}^{n-1}
\Bigl[
  -\,q^{k+1}
  \;+\;
  q^{k+1}\bigl(1 - q^{\,n-k-1}\bigr)
\Bigr]
\\[6pt]
&=\;
(1-q)(v-b)\,\sum_{k=0}^{n-1} q^k
\;+\;
\tfrac{b}{n}\,
\sum_{k=0}^{n-1}
\Bigl[
  -\,q^{k+1}
  \;+\;
  q^{k+1} - q^{k+1} q^{\,n-k-1}
\Bigr]
\\[6pt]
&=\;
(1-q)(v-b)\,\frac{1 - q^n}{1-q}
\;+\;
\tfrac{b}{n}
\sum_{k=0}^{n-1}
\bigl[-\,q^n\bigr]
\\[6pt]
&=\;
(v-b)\,\bigl(1 - q^n\bigr)
\;-\;
\tfrac{b}{n}
\;\bigl[n\,q^n\bigr]
\\[6pt]
&=\;
(v-b)\,\bigl(1 - q^n\bigr)
\;-\;
b\,q^n
\\[6pt]
&=\;
v\,\bigl(1 - q^n\bigr)
\;-\;
b.
\end{align*}

\newpage
\subsubsection*{Discrete Time Strategy}

\noindent $U_{i} \left[ b(v_i), b(v_{-i}) \right] \ = \ $ \\ 

\begin{comment}
    
\begin{align*} 
  \underbrace{ \frac{n}{\sum^n_{j=1} \left( \left[ 
        F_X \left( b(v) \right)
    \right]^{n-j} \right)} 
    }_{\text{sum of } P(\text{execute}) \text{ for all }i}
  \times \sum^n_{r=1} \Bigg( \ & 
    \underbrace{
    \left[ 
        F_X \left( b(v) \right)
    \right]^{n-r} 
    }_{P(\text{execute} \ | \ i=r)}
 \times \bigg[  
 \\ 
   & & & \left( 
   \underbrace{
    \left[1- F_X \left( b(v) \right) \right]
    }_{P(\text{success} \ | \ \text{execute})}
        \times
    \underbrace{
    \left[ v-b(v) \right]
    }_{\mathbb{E}[\text{success}]}
    \right) 
\\
    & & + & \left( 
        \underbrace{
        \left[ F_X \left( b(v) \right) \right]
        }_{P(\text{fail} \ | \ \text{execute})}
    \times 
        \underbrace{
        -
        \left[ 
        \frac{b(v)}{n} 
        -
        \left( 
            \frac{b(v)}{n}
        \left[ 1 - \left(
            F_X \left( b(v) \right) \right)^{r-1} 
        \right]
        \right)
        \right]
        }_{\max[\textit{failure cost}] \ - \ \mathbb{E}[\textit{cost reduction}] \times P_{-i}(\text{win} | -i < r)}
    \right)
\\
 \bigg] \Bigg) & & 
\end{align*}
\end{comment}

\begin{align*}
    &\underbrace{\frac{n}{\sum_{j=1}^n \left[ F_X(b(v)) \right]^{n-j}}}_{\text{sum of } P(\text{execute}) \text{ for all } i} \\
    &\quad \times \sum_{r=1}^n \Bigg( \underbrace{\left[ F_X(b(v)) \right]^{n-r}}_{P(\text{execute} \mid i=r)} \\
    &\quad \times \bigg[ \underbrace{\left[ 1 - F_X(b(v)) \right]}_{\text{$P(\text{success} \mid \text{execute})$}} \times \underbrace{\left[ v - b(v) \right]}_{\mathbb{E}[\text{success}]} \\
    &\quad + \underbrace{\left[ F_X(b(v)) \right]}_{\text{$P(\text{fail} \mid \text{execute})$}} \times \underbrace{-
        \left[ \frac{b(v)}{n} - \left( \frac{b(v)}{n} \left[ 1 - \left( F_X(b(v)) \right)^{r-1} \right] \right) \right]}_{\text{$\max[\textit{failure cost}] - \mathbb{E}[\textit{cost reduction}] \times P_{-i}(\text{win} \mid -i < r)$}} \bigg) \Bigg)
\end{align*}

\begin{align*} 
  = 
  \frac{n}{\sum^n_{j=1} \left( \left[ 
        F_X \left( b(v) \right)
    \right]^{n-j} \right)}
  \times \sum^n_{r=1} \Bigg( \ &   
\left[ 
        F_X \left( b(v) \right)
    \right]^{n-r} 
 \times \bigg[  
 \\ 
   & & & \left( 
    \left[1- F_X \left( b(v) \right) \right]
        \times 
    \left[ v-b(v) \right] 
    \right) 
\\
    & & + & \left( 
        \left[ F_X \left( b(v) \right) \right] 
    \times 
        \left[ 
        -\frac{b(v)}{n} 
        \times 
        \left( F_X \left( b(v) \right) \right)^{r-1} 
        \right]
    \right)
\\
& \bigg] \Bigg)  & & 
\end{align*}

\begin{align*} 
  = \  
  \frac{n}{\sum^n_{j=1} \left( \left[ 
        F_X \left( b(v) \right)
    \right]^{n-j} \right)}
  \times \sum^n_{r=1} \Bigg( \ &   
\\ 
    & & & \left[
        \left(
            F_X \left( b(v) \right)
    \right)^{n-r} 
 \times
    \left( [1- F_X \left( b(v) \right) \right)
\times 
    \left( v-b(v) \right) 
 \right]
 \\ 
     & & - & \left[
        \left(
            F_X \left( b(v) \right)
    \right)^{n-r} 
\times
    \left( F_X \left( b(v) \right) \right)^{r} 
 \times 
    \frac{b(v)}{n}  
 \right]
\\
 & \Bigg) 
\end{align*}

\begin{align*} 
   & = 
  \frac{
    n
    \times \sum^n_{r=1} \left( \left[
        \left(
            F_X \left( b(v) \right)
    \right)^{n-r} 
 \times
    \left( [1- F_X \left( b(v) \right) \right)
\times 
    \left( v-b(v) \right) 
    \right] 
 \ - \  
    \left[
        \left(
            F_X \left( b(v) \right)
    \right)^{n} 
 \times 
    \frac{b(v)}{n}  
 \right]
 \right)
    }{
\sum^n_{j=1} \left( \left[ 
        F_X \left( b(v) \right)
    \right]^{n-j} \right)
    } 
\\
 & = 
\frac{
    n \times  
\sum^n_{r=1} \left(   
            \left[
                F_X \left( b(v) \right)
        \right]^{n-r} 
    \times
    \left[ [1- F_X \left( b(v) \right) \right]
    \times 
    \left[ v-b(v) \right]
\right)
}{\sum^n_{j=1} \left( \left[ 
        F_X \left( b(v) \right)
    \right]^{n-j} \right)
}
\ - \ 
\frac{
    n \times \left[ 
    n 
    \ \times \ \left[
        \left(
            F_X \left( b(v) \right)
    \right)^{n} 
 \times 
    \frac{b(v)}{n}  
 \right]
\right]
}{\sum^n_{j=1} \left( \left[ 
        F_X \left( b(v) \right)
    \right]^{n-j} \right)
}
\\
 & = 
\frac{
    n 
\times
    \left[ 1- F_X \left( b(v) \right) \right]
\times 
    \left[ v-b(v) \right]
\times  
    \sum^n_{r=1} \left(   
            \left[
                F_X \left( b(v) \right)
        \right]^{n-r} 
    \right)
}{\sum^n_{j=1} \left( \left[ 
        F_X \left( b(v) \right)
    \right]^{n-j} \right)
}
\ - \ 
\frac{
    n 
    \ \times \ 
        \left(
            F_X \left( b(v) \right)
    \right)^{n} 
 \times 
    b(v)  
}{\sum^n_{j=1} \left( \left[ 
        F_X \left( b(v) \right)
    \right]^{n-j} \right)
}
\\
& = 
n 
\times\frac{
    \left(
        \left[ 1- F_X \left( b(v) \right) \right]
\times 
    \left[ v-b(v) \right]
\times  
    \sum^n_{r=1} \left[   
            \left(
                F_X \left( b(v) \right)
        \right)^{n-r} 
    \right]
    \right)
\ - \ 
    \left(
    \left[ 
        F_X \left( b(v) \right)
    \right]^{n} 
 \times 
    b(v) 
    \right)
}{\sum^n_{r=1} \left( \left[ 
        F_X \left( b(v) \right)
    \right]^{n-r} \right)
}
\\
& = 
n 
\times
\left[
\left(
    \left[ 1- F_X \left( b(v) \right) \right]
\times 
    \left[ v-b(v) \right]
\right)
\ - \
\left(
b(v)  \times
\frac{ 
    \left[ 
        F_X \left( b(v) \right)
    \right]^{n} 
}{\sum^n_{r=1} \left( \left[ 
        F_X \left( b(v) \right)
    \right]^{n-r} \right)
}
\right)
\right]
\\
& = 
n 
\times
\left[
\left(
    \left[ 1- F_X \left( b(v) \right) \right]
\times 
    \left[ v-b(v) \right]
\right)
\ - \
\left(
b(v)  \times
\frac{ 
    \left[ 
        F_X \left( b(v) \right)
    \right]^{n} 
}{
    \frac{
        1 - \left[ F_X \left( b(v) \right) \right]^{n} 
    }{
        1 - \left[ F_X \left( b(v) \right) \right]
    }
}
\right)
\right]
\\
& = 
n 
\times
\left[
\left(
    \left[ 1- F_X \left( b(v) \right) \right]
\times 
    \left[ v-b(v) \right]
\right)
\ - \
\left(
b(v)  \times
\frac{ 
    \left[ 
        F_X \left( b(v) \right)
    \right]^{n} 
    \times 
    \left(1 - \left[ F_X \left( b(v) \right) \right] \right)
}{
        1 - \left[ F_X \left( b(v) \right) \right]^{n} 
}
\right)
\right]
\\
& = 
n 
    \times
\left[ 1- F_X \left( b(v) \right) \right]
    \times
\left[
\left( v-b(v) 
\right)
\ - \
\left(
\frac{ 
    b(v)  
        \times
    \left[ 
        F_X \left( b(v) \right)
    \right]^{n} 
}{
        1 - \left[ F_X \left( b(v) \right) \right]^{n} 
}
\right)
\right]
\\
& = 
n 
    \times
\left[ 1- F_X \left( b(v) \right) \right]
    \times
\left[
\left(
\frac{ 
    \left[
    \left(
        1 - \left[ F_X \left( b(v) \right) \right]^{n} 
    \right)
    \times \left( v-b(v) 
    \right)
    \right]
    \ - \ 
    \left(
    b(v)  
        \times
    \left[ 
        F_X \left( b(v) \right)
    \right]^{n}
    \right)
}{
        1 - \left[ F_X \left( b(v) \right) \right]^{n} 
}
\right)
\right]
\\
& = 
n 
    \times
\left[ 1- F_X \left( b(v) \right) \right]
    \times
\left[
\left(
\frac{ 
    \left[
    v 
    \times
    \left(
        1 - \left[ F_X \left( b(v) \right) \right]^{n} 
    \right)
    \right]
    \ - \ 
    \left(
    b(v)  
        \times
        \left(
        1 - \left[ F_X \left( b(v) \right) \right]^{n} 
    \right)
    +
    \left[ 
        F_X \left( b(v) \right)
    \right]^{n}
    \right)
}{
        1 - \left[ F_X \left( b(v) \right) \right]^{n} 
}
\right)
\right]
\\
& = 
n 
    \times
\left[ 1- F_X \left( b(v) \right) \right]
    \times
\left[
\left(
\frac{ 
    \left[
    v 
    \times
    \left(
        1 - \left[ F_X \left( b(v) \right) \right]^{n} 
    \right)
    \right]
    \ - \ 
        b(v)  
}{
        1 - \left[ F_X \left( b(v) \right) \right]^{n} 
}
\right)
\right]
\end{align*}

\begin{align*} 
 U_{i} \left[ b(v_i), b(v_{-i}) \right] \ = \  
n 
    \times
\left[ 1- F_X \left( b(v) \right) \right]
    \times
\left[
v - 
\frac{  
        b(v)  
}{
        1 - \left[ F_X \left( b(v) \right) \right]^{n} 
}
\right]
\end{align*}

\subsubsection*{Numerical Analysis}

The equilibrium bid $b^*(v)$ is found by setting the derivative of $U_i[b(v)]$ with respect to $b(v)$ to zero. Assume \(X \sim N(v,\sigma^2)\). Letting $z = \frac{b(v)-v}{\sigma}$ and recognizing that $\Phi(z)$ and $\phi(z)$ represent the normal CDF and PDF respectively, we derive a complex condition for equilibrium:

\begin{align*}
U_i[b(v)] 
& = n 
\biggl[ 
    v (1- F_X(b(v))) 
    + \sigma f_X(b(v)) 
    - \frac{ b(v) (1 - F_X(b(v))) }{ 1 - [F_X(b(v))]^n }
\biggr]
\\ 
& = 
n \left[ 
    v(1-\Phi(z)) 
    + \sigma \frac{\phi(z)}{\sigma} 
    - \frac{b(v)(1-\Phi(z))}{1-[\Phi(z)]^n}
\right]
\\
& = 
n \left[ 
    v(1-\Phi(z)) 
    + \phi(z) 
    - \frac{b(v)(1-\Phi(z))}{1-[\Phi(z)]^n}
\right]
\end{align*}

\[
\frac{d}{db(v)}[v(1-\Phi(z))] = v(-\frac{\phi(z)}{\sigma})
\]

\begin{align*}
\frac{d}{db(v)}[\phi(z)] & = \frac{d\phi(z)}{db(v)} 
\\
& = \frac{-z\phi(z)}{\sigma}
\end{align*}

\[
Q(b(v)) = \frac{b(v)(1-\Phi(z))}{1-[\Phi(z)]^n}
\]

\[
\frac{d}{db(v)}[b(v)(1-\Phi(z))] = (1-\Phi(z)) + b(v)(-\frac{\phi(z)}{\sigma})
\]

\[
\frac{d}{db(v)}[1-(\Phi(z))^n] = -n(\Phi(z))^{n-1}\frac{\phi(z)}{\sigma}
\]

\begin{align*}
\frac{dQ}{db(v)} & = \frac{(1-[\Phi(z)]^n)\frac{d}{db(v)}[b(v)(1-\Phi(z))] - b(v)(1-\Phi(z))\frac{d}{db(v)}[1-[\Phi(z)]^n]}{[1-(\Phi(z))^n]^2}
\\
& = \frac{(1-(\Phi(z))^n)[(1-\Phi(z)) - \frac{\phi(z)}{\sigma}b(v)] - b(v)(1-\Phi(z))[-n(\Phi(z))^{n-1}\frac{\phi(z)}{\sigma}]}{[1-(\Phi(z))^n]^2}
\end{align*}

\begin{align*}
\frac{dU_i}{db(v)} & = n\Biggl[
    \frac{d}{db(v)}[v(1-\Phi(z))] 
    + \frac{d}{db(v)}[\phi(z)]
    - \frac{dQ}{db(v)}
\Biggr]
\\
& = n\Biggl[
    -\frac{v\phi(z)}{\sigma}
    - \frac{z\phi(z)}{\sigma}
    - \frac{dQ}{db(v)}
\Biggr]
\\
& = n\left[ 
    -\frac{v \cdot\Phi(z)}{\sigma} 
    - z \cdot \Phi(z) 
    + \frac{1 - \frac{\phi(z) - b(v) \cdot \Phi(z)}{\sigma}}{1-(\frac{ \phi(z)}{\sigma})^n} 
    + \frac{(1-\frac{\phi(z)}{\sigma})b(v)n(\frac{\phi(z)}{\sigma})^{n-1}\Phi(z)/\sigma}{(1-(\frac{\phi(z)}{\sigma})^n)^2}
\right]
\end{align*}

\end{document}